\documentclass[showpacs,preprintnumbers,amsmath,amssymb,prb,preprint]{revtex4}
\usepackage{graphicx}
\usepackage{dcolumn}
\usepackage{bm}
\begin{document}
\preprint{}
\title{Complex electronic states in double layered ruthenates
(Sr$_{1-x}$Ca$_{x}$)$_{3}$Ru$_{2}$O$_{7}$}
\author{Zhe Qu$^{1}$, Jin Peng$^{1}$, Tijiang Liu$^{1}$, David
Fobes$^{1}$, Leonard Spinu$^{2}$, and Zhiqiang Mao$^{1}$}\email{zmao@tulane.edu}
\affiliation{1. Department of Physics and Engineering Physics, Tulane University, New Orleans,
Louisiana 70118, USA. \\ 2. Advanced Material Research Institute and
Physics Department, University of New Orleans, Louisiana 70148,
USA.}

\date{\today}
\begin{abstract}
The magnetic ground state of (Sr$_{1-x}$Ca$_x$)$_3$Ru$_2$O$_7$ (0
$\leq x \leq$ 1) is complex, ranging from an itinerant metamagnetic
state (0 $\leq x <$ 0.08), to an unusual heavy-mass, nearly
ferromagnetic (FM) state (0.08 $< x <$ 0.4), and finally to an
antiferromagnetic (AFM) state (0.4 $\leq x \leq$ 1). In this report
we elucidate the electronic properties for these magnetic states,
and show that the electronic and magnetic properties are strongly
coupled in this system. The electronic ground state evolves from an
AFM quasi-two-dimensional metal for $x =$ 1.0, to an Anderson
localized state for $0.4 \leq x < 1.0$ (the AFM region). When the
magnetic state undergoes a transition from the AFM to the nearly FM
state, the electronic ground state switches to a weakly localized
state induced by magnetic scattering for $0.25 \leq x < 0.4$, and
then to a magnetic metallic state with the in-plane resistivity
$\rho_{ab} \propto T^\alpha$ ($\alpha >$ 2) for $0.08 < x < 0.25$.
The system eventually transforms into a Fermi liquid ground state
when the magnetic ground state enters the itinerant metamagnetic
state for $x < 0.08$. When $x$ approaches the critical composition
($x \sim$ 0.08), the Fermi liquid temperature is suppressed to zero
Kelvin, and non-Fermi liquid behavior is observed. These results
demonstrate the strong interplay between charge and spin degrees of
freedom in the double layered ruthenates.
\end{abstract}
 \maketitle

\section{Introduction}

The Ruddlesden-Popper series of perovskite ruthenates
(Sr,Ca)$_{n+1}$Ru$_{n}$O$_{3n+1}$ have attracted significant
attention since they exhibit a wide range of unique electronic and
magnetic states. The richness of states in ruthenates is epitomized
by unconventional spin-triplet superconductivity
\cite{Sr214SC,Sr214NMRnature,Sr214phasesentiveScience},
antiferromagnetic (AFM) Mott insulating behavior
\cite{SrCa214NakatsujiPRL1,SrCa214NakatsujiPRL2,SrCa214EMIT}, an
electronic nematic phase \cite{Sr327Nematic}, itinerant magnetism
\cite{Sr113GS,SrCa113}, and orbital ordered states
\cite{Ca214OOPRL1,Ca214OOPRL2,Ca327OOCooperPRL}. These states all
occur in close proximity and provide a rare opportunity to tune the
system between various states using non-thermal control parameters
such as chemical composition, pressure, and magnetic field. Tuning
of non-thermal parameters often results in interesting exotic
properties.

The double layered ruthenates
(Sr$_{1-x}$Ca$_{x}$)$_{3}$Ru$_{2}$O$_{7}$  provide a typical
example. The properties of the end members in this series are
dramatically different. The $x =$ 0 member, Sr$_{3}$Ru$_{2}$O$_{7}$,
is an enhanced paramagnet showing an itinerant metamagnetic
transition \cite{Sr327GS}. Its electronic properties under magnetic
fields remarkably depend on field orientation
\cite{Sr327Nematic,Sr327QC,Sr327PerryPRL}. Magnetic field applied
along the $c$-axis induces an electronic nematic phase near a
metamagnetic quantum critical end point
\cite{Sr327QC,Sr327PerryPRL,Sr327Nematic}. In contrast,
Ca$_{3}$Ru$_{2}$O$_{7}$ ($x =$ 1) is AFM with N\'{e}el temperature
$T_{\mathrm{N}} =$ 56 K \cite{Ca327AFM,Ca327Q2D} and exhibits giant
magnetoresistance attributed to a bulk spin-valve effect
\cite{Ca327GMR,Ca327BSV,Ca327YoshidaNeutronPRB,Ca327elastic}. Recent
studies on floating-zone grown high quality single crystals of
(Sr$_{1-x}$Ca$_{x}$)$_{3}$Ru$_{2}$O$_{7}$ revealed rich exotic
magnetic properties \cite{SrCa327HMNF,SrCa327JPSJ}. With Ca
substitution for Sr, the system evolves from an itinerant
metamagnetic state (0 $\leq x <$ 0.08) to an unusual heavy-mass,
nearly ferromagnetic (FM) state with an extremely large Wilson ratio
$R_{\mathrm{w}}$ (0.08 $< x <$ 0.4). $R_{\mathrm{w}}$ is $\sim$ 10
for Sr$_3$Ru$_2$O$_7$ \cite{Sr327GS}; it increases to a maximum of
$\sim$ 700 near $x = 0.2$ \cite{SrCa327HMNF}. The nearly FM state
does not develop a long-range FM order despite considerably strong
FM correlations manifested by the large Wilson ratio, but instead
freezes in a cluster-spin-glass (CSG) phase at low temperatures
\cite{SrCa327HMNF,SrCa327JPSJ}. The system eventually switches to a
long range AFM state for $0.4 \leq x \leq 1$
\cite{SrCa327HMNF,SrCa327JPSJ}, in which the magnetic easy axis
changes continuously from the $c$-axis to the $ab$-plane with
increasing Ca content \cite{SrCa327JPSJ}.

To better understand these complex magnetic phase transitions,
information on the electronic states involved in these transitions
is needed. The end members of
(Sr$_{1-x}$Ca$_{x}$)$_{3}$Ru$_{2}$O$_{7}$ have been shown to exhibit
distinctly different electronic states; while
Sr$_{3}$Ru$_{2}$O$_{7}$ exhibits Fermi liquid behavior
\cite{Sr327GS}, Ca$_{3}$Ru$_{2}$O$_{7}$ undergoes a metal-insulator
transition (MIT) at 48 K \cite{Ca327AFM,Ca327Q2D}, which has been
suggested to be caused by the opening of a pseudogap associated with
a density wave instability \cite{Ca327ARPES,Ca327pseudogap}, and
enters a quasi-two-dimensional (2D) metallic ground state at low
temperatures \cite{Ca327Q2D}. It is particularly interesting to
investigate the evolution of the electronic states between these two
end members and how the states are coupled to the magnetic states.

Here, we report a systematic study on the electronic states of the
(Sr$_{1-x}$Ca$_{x}$)$_{3}$Ru$_{2}$O$_{7}$ solid solution by means of
transport and specific heat measurements. We find that the
electronic states are complex and strongly coupled with the magnetic
states in this system. In the AFM state with 0.4 $\leq x <$ 1.0, the
electronic ground state of the system behaves as an Anderson
localized state. In the nearly FM region, however, the system shows
different electronic properties: a weakly localized state induced by
magnetic scattering is observed for $0.25 \leq x < 0.4$ and a
magnetic metallic state with the resistivity $\rho \propto T^\alpha$
($\alpha
>$ 2) occurs for $0.08 < x <0.25$. When the nearly FM state
transforms to a metamagnetic state for $x <$ 0.08, the electronic
state changes to a Fermi liquid ground state. The Fermi liquid
temperature is suppressed to zero Kelvin when $x$ approaches the
critical value 0.08. Non-Fermi liquid behavior is also observed in
thermodynamic properties near this critical composition.

\section{Experiment}

We have succeeded in growing high quality single crystal samples of
(Sr$_{1-x}$Ca$_{x}$)$_{3}$Ru$_{2}$O$_{7}$ in the whole range of $x$
using the floating-zone technique \cite{SrCa327HMNF}. All crystals
selected for measurements were characterized carefully by x-ray
diffraction and SQUID magnetometer. Since SQUID has an extremely
high sensitivity to ferromagnetic materials, it guarantees that the
selected samples do not contain any FM impurity phases such as
(Sr,Ca)$_{4}$Ru$_{3}$O$_{10}$ \cite{SrCa4310} and (Sr,Ca)RuO$_{3}$
\cite{SrCa113}. SQUID was also used for magnetization measurements
on selected samples. Specific heat measurements were performed using
a thermal relaxation method in a Quantum Design PPMS. The electrical
transport measurements were carried out in a He$^{3}$ cryostat with
a base temperature of 0.3 K, using a standard four-probe technique.
The crystallographic axes of the samples selected for transport
measurements were identified using X-ray Laue diffraction. The
electrical current was applied along in-plane Ru-O-Ru bond direction
for in-plane resistivity $\rho_{ab}$ measurements and along the c-axis
for out-of-plane resistivity $\rho_{c}$ measurements.

\section{Results}

\begin{figure}
\includegraphics[angle=0]{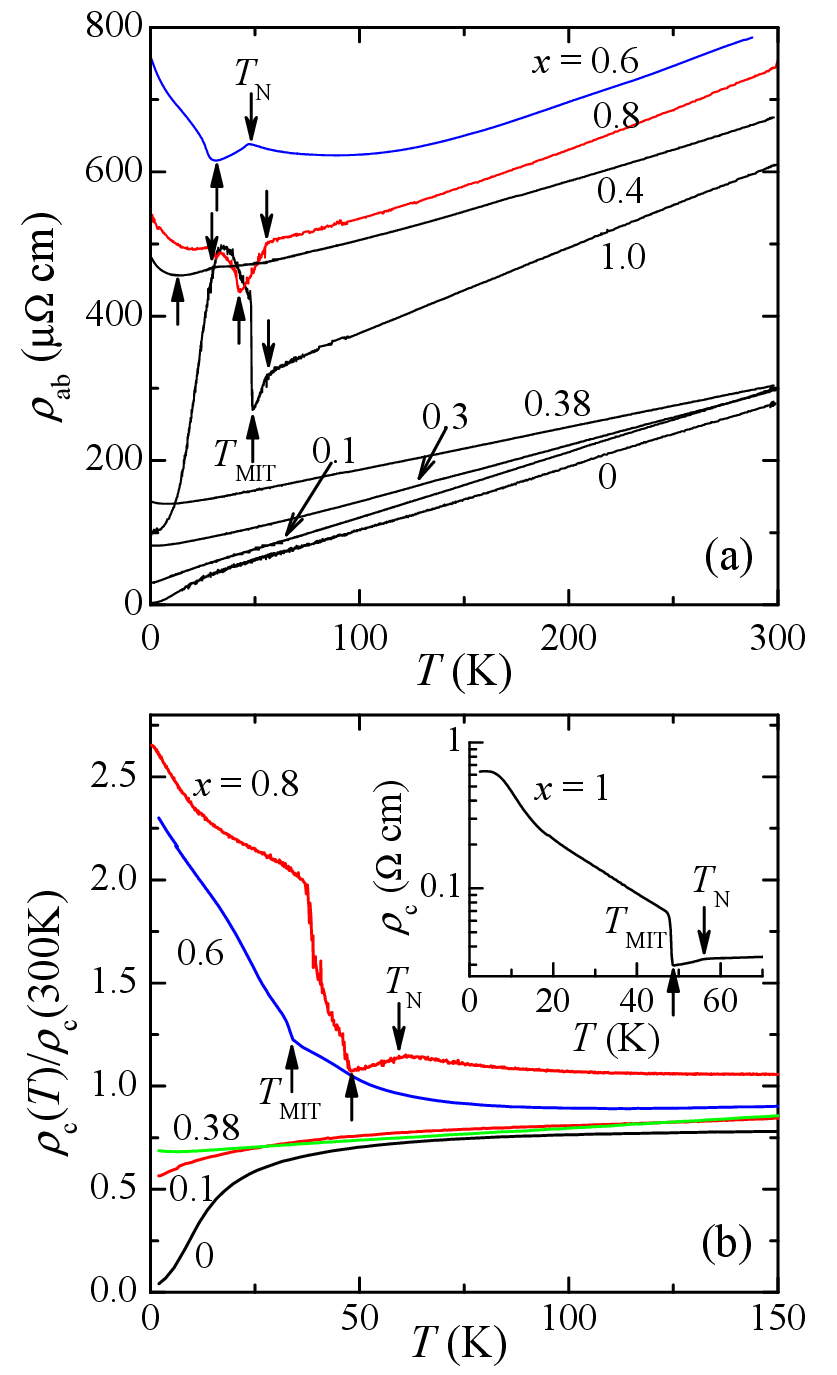}
\caption{(color online) (a) Temperature dependence of in-plane
resistivity $\rho_{ab}$($T$) for double layered ruthenates
(Sr$_{1-x}$Ca$_{x}$)$_{3}$Ru$_{2}$O$_{7}$. The downward arrow
denotes the N\'{e}el temperature $T_{\mathrm{N}}$ and the upward
arrow marks the metal-insulator transition temperature
$T_{\mathrm{MIT}}$. (b) Temperature dependence of out-of-plane
resistivity $\rho_{c}$($T$) (normalized to its 300 K value) for
typical compositions. Inset shows $\rho_{c}$ as a function of
temperature for Ca$_3$Ru$_2$O$_7$ ($x =$ 1). }\label{fig:rhoabT}
\end{figure}

Figure \ref{fig:rhoabT} shows the temperature dependence of
resistivity, $\rho_{ab}$($T$) and $\rho_{c}$($T$), for typical
compositions of the (Sr$_{1-x}$Ca$_{x}$)$_{3}$Ru$_{2}$O$_{7}$ solid
solution series. Like other layered ruthenates with cylindrical
Fermi Surfaces (FS)
\cite{Sr214FS,Sr214FL,Sr214ES,Sr214QO,Sr214BS,Sr214RE},
(Sr$_{1-x}$Ca$_{x}$)$_{3}$Ru$_{2}$O$_{7}$ displays remarkable
anisotropy between $\rho_{ab}$ and $\rho_{c}$. $\rho_{c}$ is much
higher than $\rho_{ab}$ throughout the entire series. For
Ca$_{3}$Ru$_{2}$O$_{7}$ ($x = 1.0$), both $\rho_{ab}$ and $\rho_{c}$
exhibit anomalies at temperatures corresponding to $T_{\mathrm{N}}$
and $T_{\mathrm{MIT}}$; they show metallic behavior immediately
below $T_{\mathrm{N}} = 56$ K, but a discontinuous increase at
$T_{\mathrm{MIT}} = 48$ K. Below $T_{\mathrm{MIT}}$, $\rho_{ab}$
first increases with decreasing temperature, then exhibits metallic
behavior once again at low temperatures. These characteristics are
consistent with previous results obtained using floating-zone grown
crystals \cite{Ca327Q2D}. Recent studies demonstrate that the low
temperature metallic behavior originates from small FS pockets which
survive below $T_{\mathrm{MIT}}$ \cite{Ca327ARPES}. With Sr
substitution for Ca, $T_{\mathrm{N}}$ and $T_{\mathrm{MIT}}$, as
determined from $\rho_{ab}$, systematically shift to lower
temperatures for 0.4 $\leq x <$ 1 and are consistent with those
obtained early from magnetization measurements \cite{SrCa327HMNF}.
The system still remains an AFM metallic state between
$T_{\mathrm{N}}$ and $T_{\mathrm{MIT}}$ in this composition range.
However in contrast to the first-order-like MIT observed in
Ca$_{3}$Ru$_{2}$O$_{7}$, the MIT in doped samples occurs in a
continuous fashion. Below $T_{\mathrm{MIT}}$, both $\rho_{ab}$ and
$\rho_c$ exhibit non-metallic behavior down to 2K ($d\rho/dT < 0$),
consistent with the results reported by Iwata \textit{et al.},
\cite{SrCa327JPSJ}. We also note that $\rho_c$ show anomalies at
$T_{\mathrm{N}}$ and $T_{\mathrm{MIT}}$ for 0.6 $< x \leq$ 1, but
only at $T_{\mathrm{MIT}}$ for 0.4 $\leq x \leq$ 0.6. These results
suggest that partial Sr substitution for Ca $(0.4 \leq x < 1)$ leads
the electronic ground state to transform from a quasi 2D metallic
state in Ca$_{3}$Ru$_{2}$O$_{7}$ to a non-metallic state. This
non-metallic state can be ascribed to Anderson localization as
discussed below.

\begin{figure}
\includegraphics[angle=-90,scale=1]{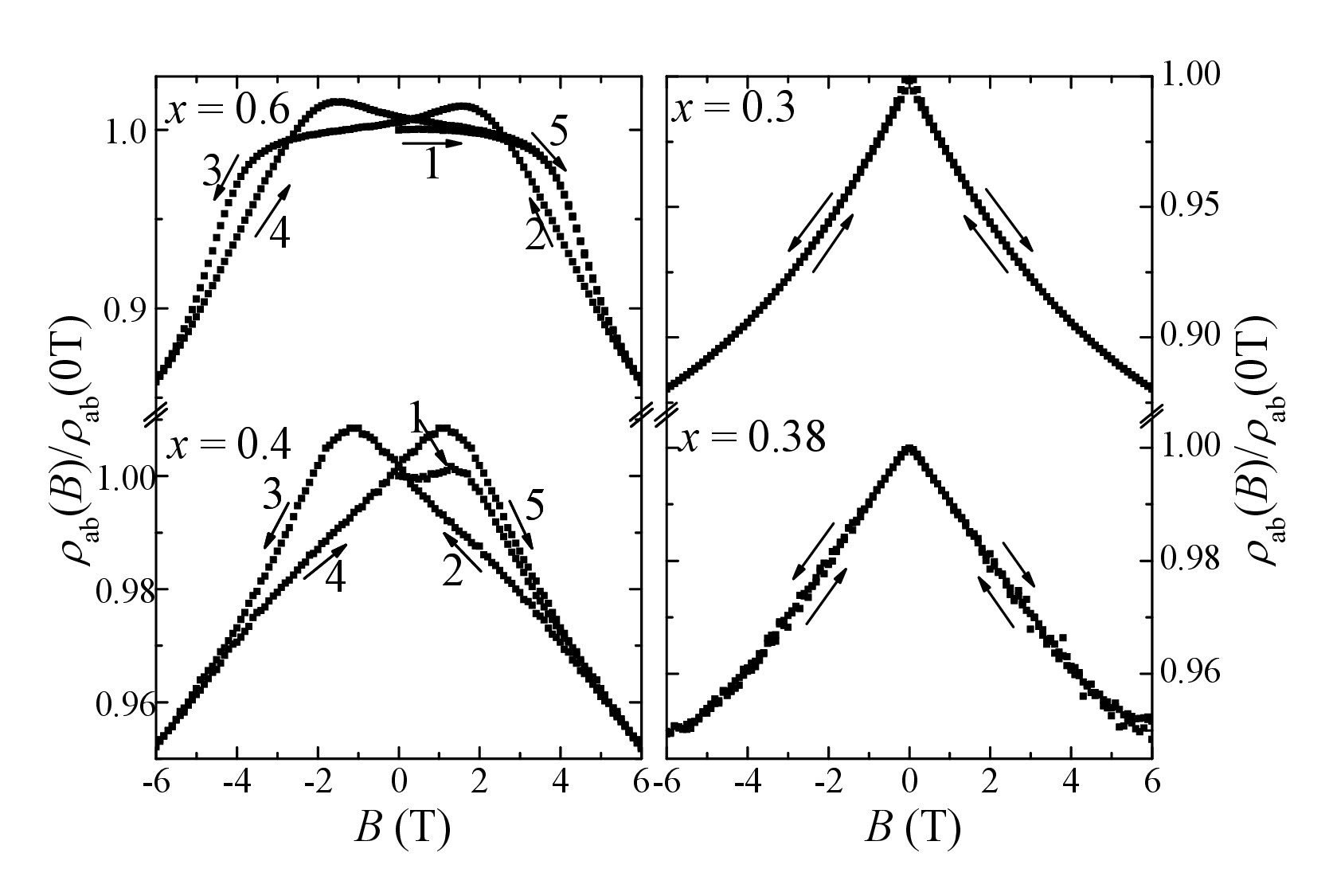}
\caption{In-plane resistivity (normalized to its 0 T values) at 0.3
K versus field for (Sr$_{1-x}$Ca$_{x}$)$_{3}$Ru$_{2}$O$_{7}$. The
magnetic field is applied parallel to the electrical
current.}\label{fig:MR}
\end{figure}

\begin{figure}
\includegraphics[angle=0]{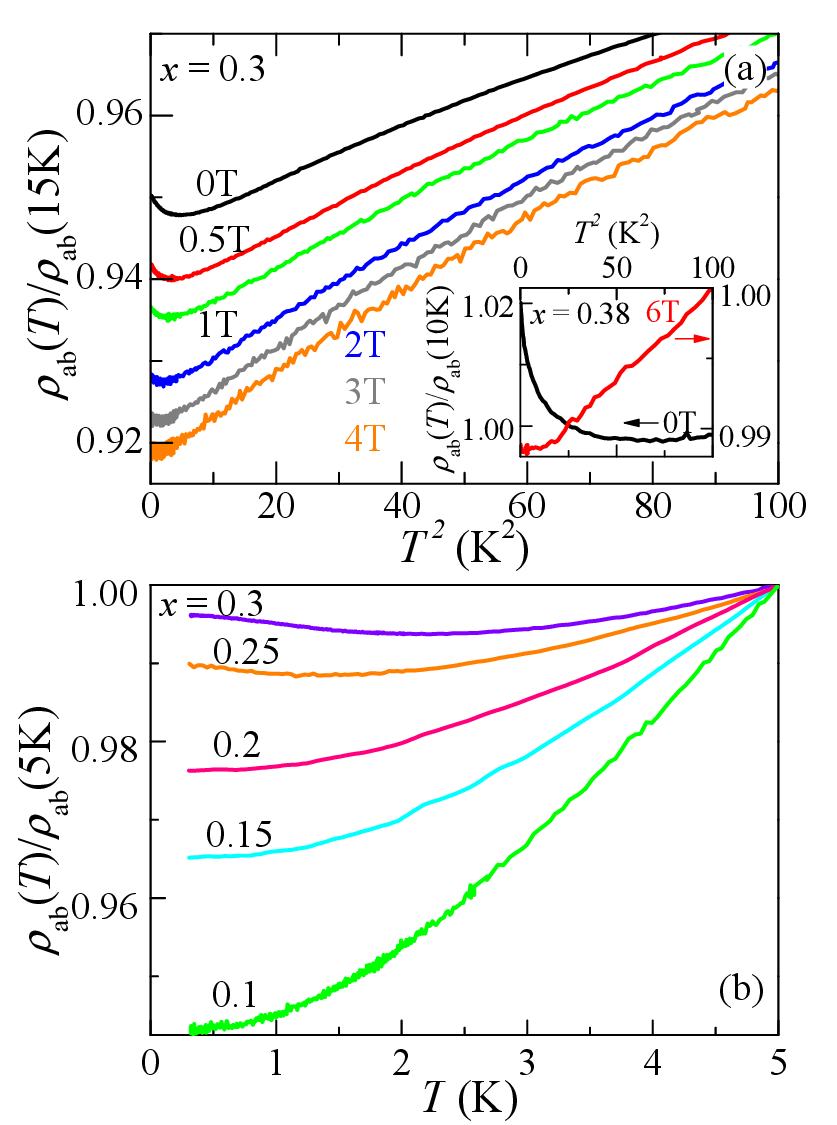}
\caption{(color online) (a) In-plane resistivity (normalized to its
15 K value) plotted versus $T^2$ under a range of applied fields,
for the $x = 0.3$ sample. Inset: In-plane resistivity (normalized
to its 10 K value) plotted versus $T^2$ under zero field and 6 T for
the $x = 0.38$ sample. (b)Temperature dependence of the in-plane
resistivity for several samples with $0.1 \leq x \leq 0.3$,
normalized to their values at 5 K.}\label{fig:RTB}
\end{figure}

\begin{figure}
\includegraphics[angle=-90]{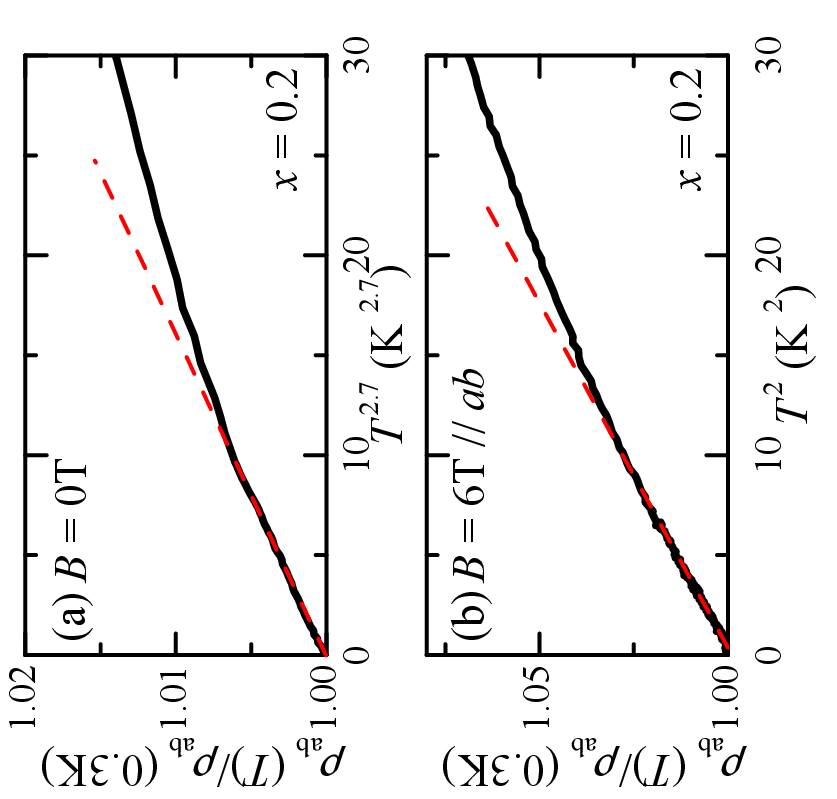}
\caption{(color online) (a) In-plane resistivity (normalized to its
value at 0.3 K) plotted versus $T^{2.7}$ under zero field for the $x
= 0.2$ sample. (b) In-plane resistivity (normalized to its value at
0.3 K) plotted versus $T^2$ under 6 T for the $x=0.2$
sample.}\label{fig:powerlaw}
\end{figure}

\begin{figure}
\includegraphics[angle=-90]{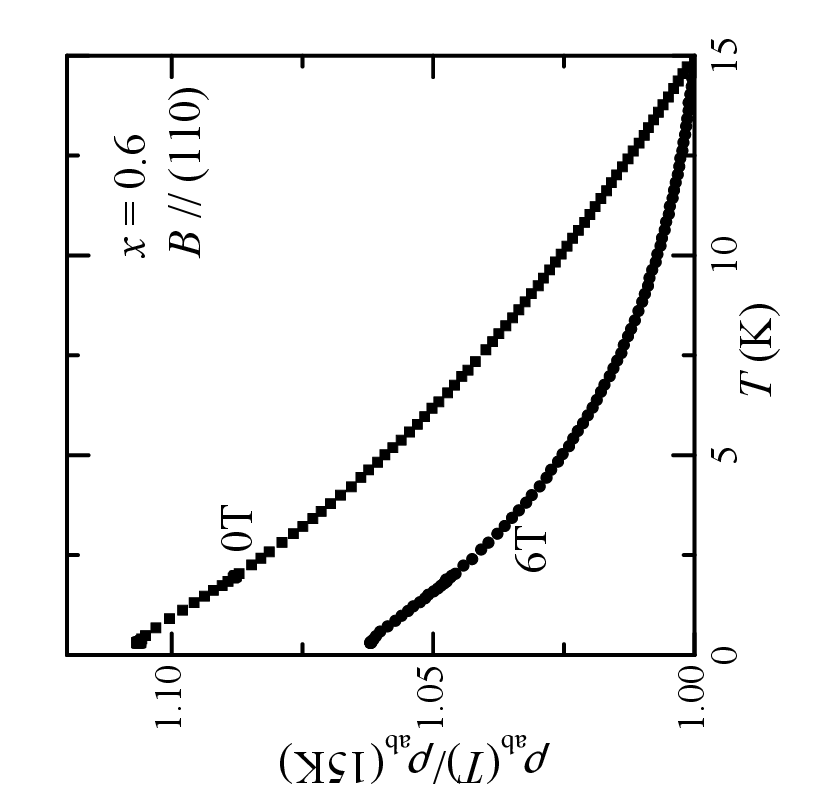}
\caption{Temperature dependence of the in-plane resistivity
(normalized to its 15K value) under zero field and 6 T for the
$x=0.6$ sample. The magnetic field is applied parallel to the
electrical current.}\label{fig:suppress}
\end{figure}

When $x < 0.4$, a magnetic phase transition from the AFM to the
heavy-mass, nearly FM state occurs \cite{SrCa327HMNF,SrCa327JPSJ}.
This transition is also probed in magnetoresistivity measurements,
as shown in Fig. \ref{fig:MR}. For $x \geq 0.4$, a metamagnetic
transition with clear hysteresis occurs, ascribed to the spin
flip/flop transition of the AFM state. The metamagnetic transition
field increases with increasing $x$ and exceeds 6 T for $x \geq$
0.8. In these measurements the magnetic field is applied along the
in-plane Ru-O-Ru direction, \textit{i.e.} the (110) direction in the
$Bb2_{1}m$ space group, which is not the easy axis of magnetization
\cite{SrCa327StrM,Ca327Q2D,Ca327HA}. For $x = 0.3$ and 0.38,
however, a negative magnetoresistance without hysteresis is
observed, consistent with the expected result for the nearly FM
state \cite{SrCa327HMNF,SrCa327JPSJ}. Accompanied with the magnetic
phase transition, the electronic state changes drastically. As seen
in Fig. \ref{fig:rhoabT}, the resistivity of the nearly FM state in
0.08 $< x <$ 0.4 is much smaller than that of the AFM state in 0.4
$\leq x <$1. Its temperature dependence show two different
behaviors: for 0.25 $\leq x <$ 0.4 $\rho_{ab}$ exhibits a small
upturn at low temperature (see Fig. \ref{fig:RTB}b), suggesting a
weakly localized state, while for $0.08 < x < 0.25$, $\rho_{ab}$
shows metallic behavior in the whole temperature range and can be
fitted to $\rho \propto T^\alpha$ ($\alpha >$ 2) at low temperatures
(\textit{e.g.} $\alpha = 2.7$ for $x = 0.2$ (see Fig.
\ref{fig:powerlaw}a)). Interestingly, both behaviors are sensitive
to magnetic field, a moderate magnetic field can tune them to
quasi-quadratic temperature dependences, as show in Fig.
\ref{fig:RTB}a and Fig. \ref{fig:powerlaw}b, suggesting that they
are of magnetic origin, which will be discussed below. In contrast,
the non-metallic state in the AFM region with $0.4 \leq x < 1$ is
more robust, less sensitive to magnetic field as shown in Fig.
\ref{fig:suppress}.

\begin{figure}
\includegraphics[angle=0]{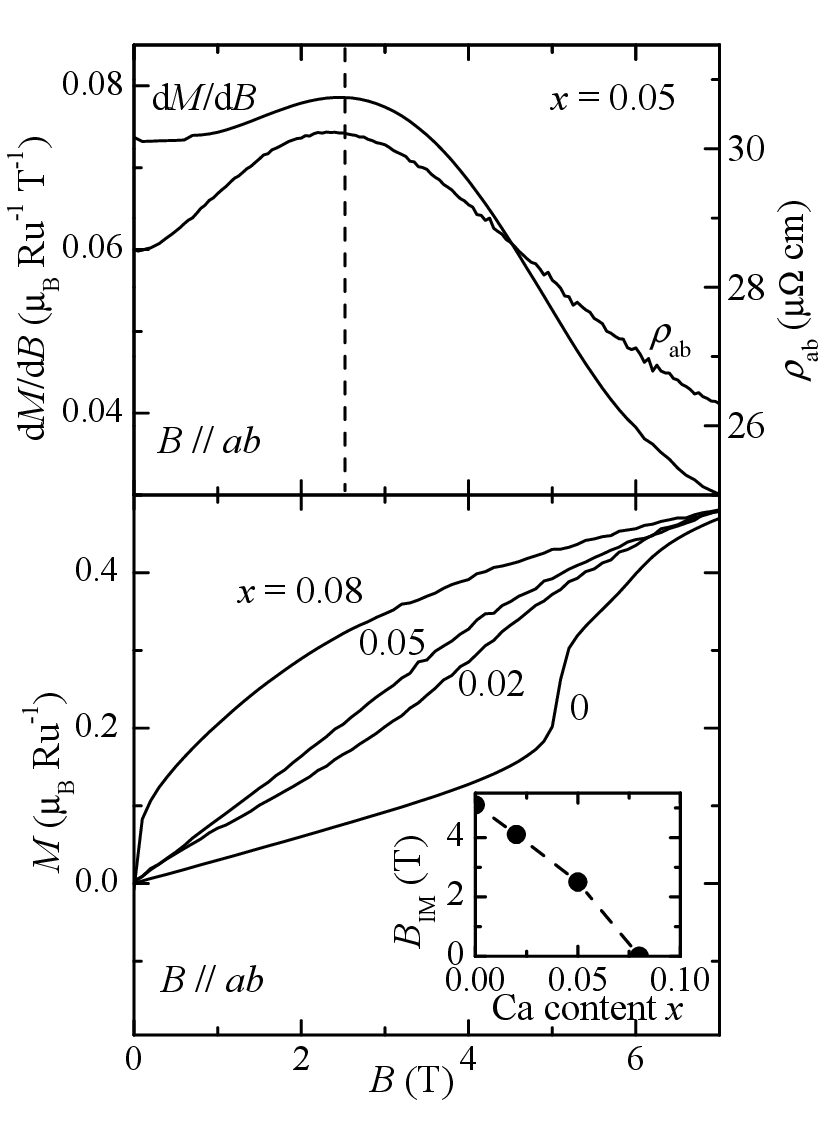}
\caption{Upper panel: in-plane resistivity and the first derivative
of the magnetization $dM/dB$ as a function of magnetic field at $T =
2$ K for the $x = 0.05$ sample. The dash line marks the metamagnetic
transition field $B_{\mathrm{IM}}$. Lower panel: magnetization as a
function of magnetic field at 2 K for the samples with $x = 0.08,
0.05, 0.02$ and 0. Inset: the evolution of the
metamagnetic transition field as a function of the Ca
content.}\label{fig:meta}
\end{figure}

\begin{figure}
\includegraphics[angle=0]{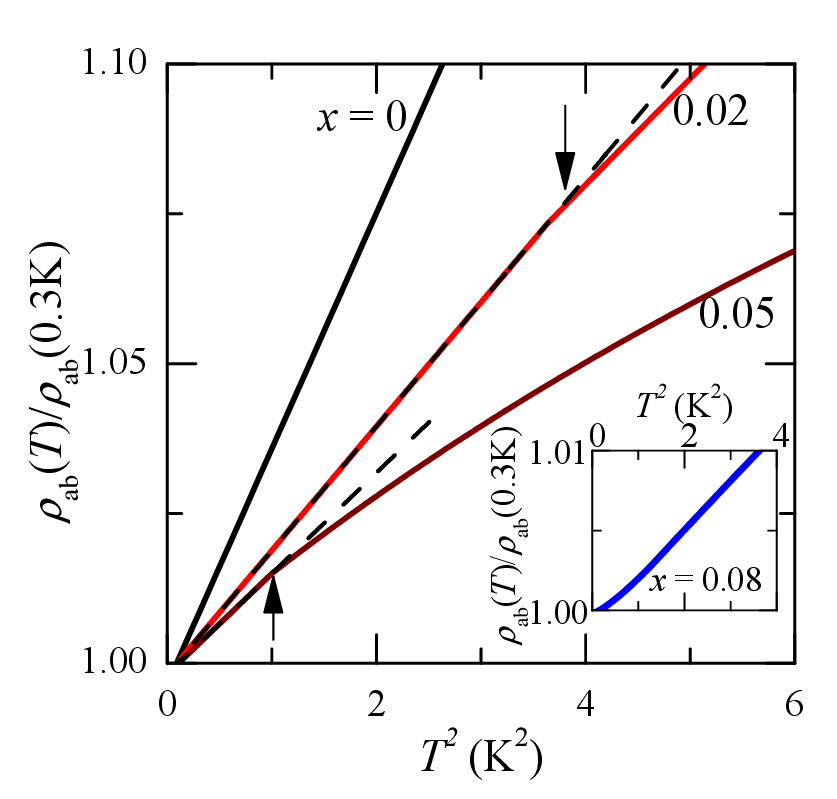}
\caption{(color online) In-plane resistivity (normalized to its 0.3
K value) plotted versus $T^2$ for several samples with various Ca
contents. Arrows indicate the temperature where the curve deviates
from linearity, \textit{i.e.} the Fermi liquid temperature
$T_{\mathrm{FL}}$. Inset shows the data for the $x =$ 0.08
sample.}\label{fig:FL}
\end{figure}

When the Ca content is decreased below $x = 0.08$, the system
transforms into a metamagnetic Fermi liquid ground state. As shown
in the upper panel of the Fig. \ref{fig:meta}, the metamagnetic
transition can be identified in both resistivity and magnetization
for this composition range. In contrast to the spin flip/flop
induced metamagnetic transition seen in the samples with $x \geq
0.4$, the metamagnetic transition observed here is reversible
between upward and downward field sweeps and this transition is
generally interpreted as a field-induced Stoner transition
\cite{Sr327PerryPRL}. The metamagnetic transition field
$B_{\mathrm{IM}}$ depends sensitively on the Ca content; for $x$ =
0, $B_{\mathrm{IM}} \sim $5.1 T, consistent with the previous result
\cite{Sr327PerryPRL}; $B_{\mathrm{IM}}$ decreases rapidly as $x$
increases, down to zero as $x$ approaching the critical value 0.08
(see the inset to the lower panel of Fig. \ref{fig:meta}). Occurring
along with the metamagnetism, as shown in Fig. \ref{fig:FL}, the
resistivity exhibits a quadratic temperature dependence below a
characteristic temperature $T_{\mathrm{FL}}$, indicating a Fermi
liquid ground state. The Fermi liquid temperature $T_{\mathrm{FL}}$
is $\sim$ 10 K for $x = 0$; it decreases rapidly to $\sim 1.95$ K
for $x = 0.02$, and is eventually suppressed to zero near $x =
0.08$.

From the above results for various composition ranges, we have
constructed an electronic phase diagram for
(Sr$_{1-x}$Ca$_{x}$)$_{3}$Ru$_{2}$O$_{7}$, and plotted it together
with the magnetic phase diagram we obtained earlier
\cite{SrCa327HMNF}, as shown in the upper panel of Fig.
\ref{fig:ps}. To summarize, Ca$_{3}$Ru$_{2}$O$_{7}$ shows an AFM
metallic state below $T_{\mathrm{N}} = 56$ K and then experiences a
metal-insulator transition at $T_{\mathrm{MIT}} = 48$ K, eventually
evolving into a quasi two-dimensional metallic ground state at low
temperatures. With Sr substitution for Ca, the metallic state
remains between $T_{\mathrm{N}}$ and $T_{\mathrm{MIT}}$ ($0.4 \leq x
<1$, Region I) and yields to an Anderson localized state caused by
disorders as the temperature is decreased below $T_{\mathrm{MIT}}$
($0.4 \leq x <1$, Region II). When $x < 0.4$, the magnetic ground
state switches from the AFM to the nearly FM state that freezes to a
CSG phase at low temperatures \cite{SrCa327HMNF,SrCa327JPSJ}.
Accompanying this magnetic phase transition is an electronic ground
state transition from the Anderson localized state to a
weakly localized state induced by magnetic scattering ($0.25 \leq x
< 0.4$, Region III). Further decrease of \textit{x} leads to the
presence of a magnetic metallic state ($0.08 < x < 0.25$, Region
IV). When $x < 0.08$, the system enters an itinerant metamagnetic
Fermi liquid ground state (Region V).

\section{Discussion}

First we will discuss the mechanism behind the MIT between Region I
and II. In a strongly correlated electron system, MITs can generally
be separated into two categories according to their driving forces
\cite{MITreview}, \textit{i.e.} a Mott transition due to the
correlation effect of electrons , and an Anderson transition due to
disorder. In a typical Mott transition, on-site Coulomb interactions
create strongly renormalized quasiparticle states and open a gap at
the Fermi level \cite{MITMottreview}. Therefore, there should be no
density of states at Fermi level, resulting in an electronic
specific heat equal to zero for a Mott-type insulator. This is well
known and has been observed in many materials, \textit{e.g.} single
layered ruthenate Ca$_{2}$RuO$_{4}$ \cite{Ca214Mott}.

\begin{figure}
\includegraphics[angle=0]{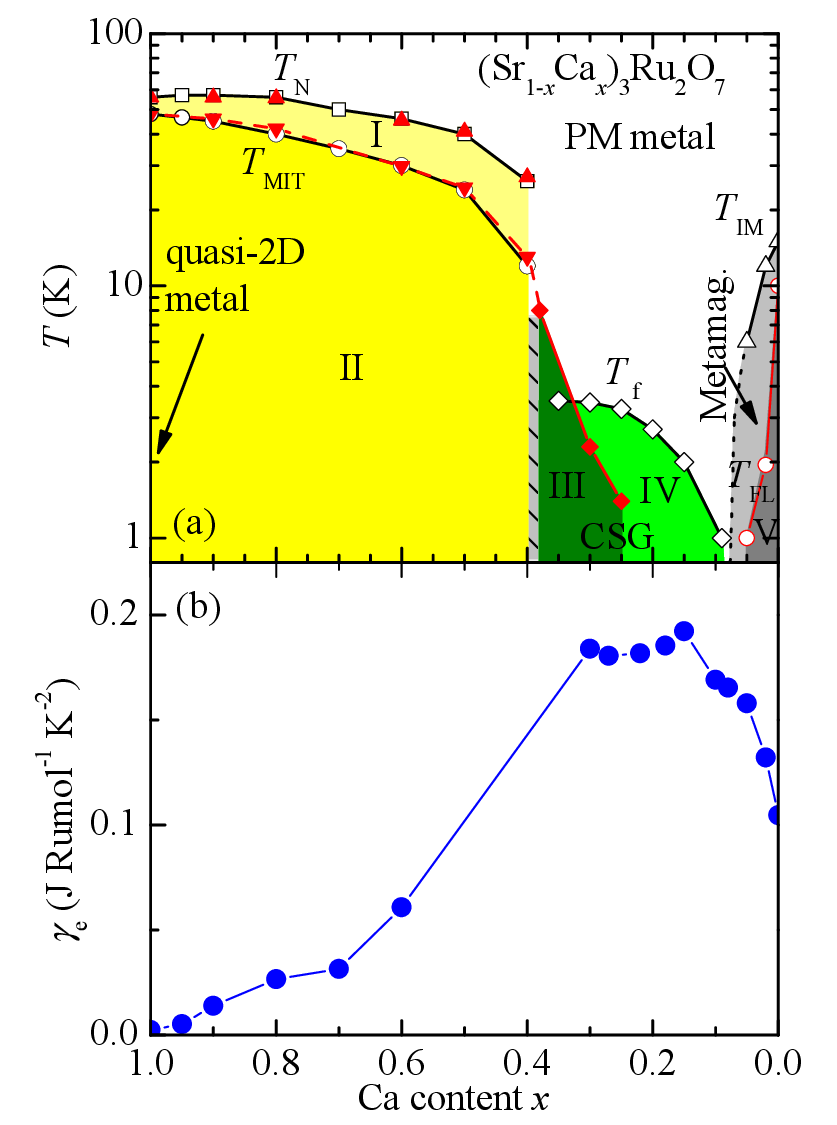}
\caption{(color online) (a) The electronic phase diagram of the
double layered ruthenates (Sr$_{1-x}$Ca$_{x}$)$_{3}$Ru$_{2}$O$_{7}$.
$T_{\mathrm{N}}$: N\'{e}el temperature; $T_{\mathrm{MIT}}$:
metal-insulator transition temperature. The open squares $\square$
and open circles $\bigcirc$ represent $T_{\mathrm{N}}$ and
$T_{\mathrm{MIT}}$ determined from magnetization measurements
\cite{SrCa327HMNF}, and the filled upward and downward triangles
($\blacktriangle$ and $\blacktriangledown$) represent those
determined from $\rho_{ab}$ measurements. $T_{\mathrm{f}}$ is the
freezing temperature for the cluster spin glass (CSG) phase
\cite{SrCa327HMNF}. The filled diamond $\blacklozenge$ is the
temperature below which the electronic state becomes weakly
localized due to magnetic scattering. $T_{\mathrm{IM}}$: the
characteristic temperature below which an itinerant metamagnetic
phase transition occurs \cite{SrCa327HMNF}. $T_{\mathrm{FL}}$ is the
Fermi liquid temperature. Region I: the AFM metallic state. Region
II: the AFM Anderson localized state. Region III: the weakly
localized state induced by magnetic scattering. Region IV: the
magnetic metallic state (see text). Region V: the metamagnetic Fermi
liquid state. (b) The electronic specific heat coefficient
$\gamma_{e}$ as a function of Ca content \textit{x}.}\label{fig:ps}
\end{figure}

\begin{figure}
\includegraphics[angle=-90]{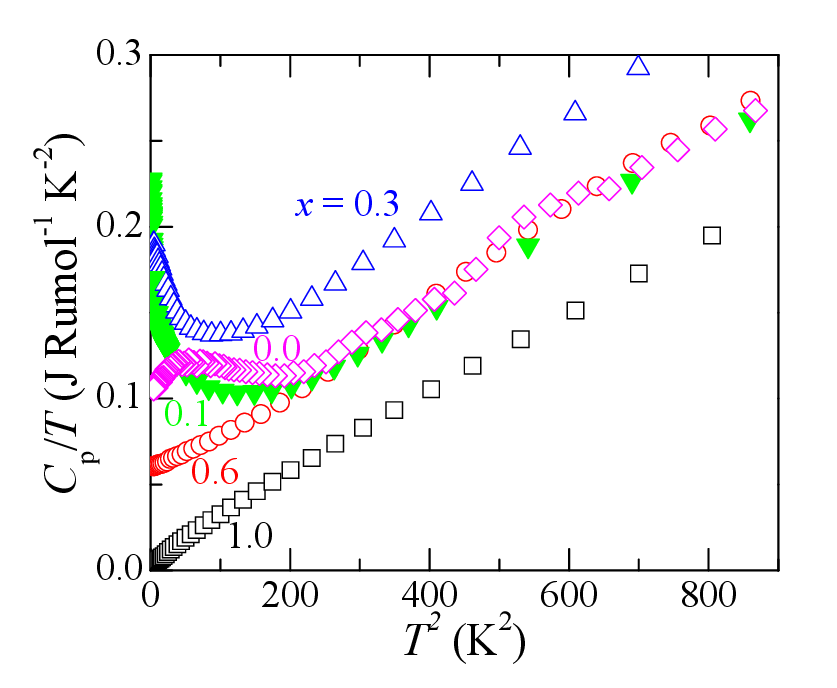}
\caption{(color online) Specific heat divided by temperature
$C/T$ plotted versus $T^2$ for typical samples.}\label{fig:Cp}
\end{figure}

However, the MIT observed in
(Sr$_{1-x}$Ca$_{x}$)$_{3}$Ru$_{2}$O$_{7}$ for $0.4 \leq x \leq 1.0$
should not be attributed to a typical Mott transition.
Ca$_{3}$Ru$_{2}$O$_{7}$ was initially thought to be a Mott-like
system with a small charge gap $\sim 0.1$ eV
\cite{Ca327AFM,Ca327CaoPRB62998,Ca327Raman}. But later, an
angle-resolved photoemission spectroscopy (ARPES) study revealed
that small, non-nesting Fermi pockets survive even at lowest
achievable temperatures \cite{Ca327ARPES}, leading to a small, but
nevertheless, non-zero electronic specific heat coefficient
($\gamma_{\mathrm{e}} = $1.7 mJ/Rumol K$^2$) and a quasi-2D metallic
transport behavior \cite{Ca327Q2D}. In order to examine how
$\gamma_{\mathrm{e}}$ changes with Sr substitution for Ca, we
measured the specific heat of
(Sr$_{1-x}$Ca$_{x}$)$_{3}$Ru$_{2}$O$_{7}$; the data is shown in Fig.
\ref{fig:Cp}. Since the phonon term is proportional to $T^{3}$, the
phonon contribution can be evaluated by a linear fitting of $C/T$
vs. $T^{2}$. The electronic specific heat is obtained by subtracting
the phonon contribution from the total specific heat. From these
analyses, we obtained $\gamma_{\mathrm{e}}$ for most of our samples,
as shown in Fig. \ref{fig:ps}(b). We find that $\gamma_{\mathrm{e}}$
is indeed small for Ca$_3$Ru$_2$O$_7$, consistent with the previous
result \cite{Ca327Q2D}; the Sr substitution for Ca gradually
increases $\gamma_{\mathrm{e}}$, suggesting that the size of FS
increases. These results are clearly not expected for a typical Mott
insulator, therefore the non-metallic state observed in Region II
cannot be attributed to a Mott insulating state.

Since Sr substitution for Ca introduces disorders to the system,
disorder-driven Anderson localization effect should be considered.
In general, in the presence of Anderson localization, there exists a
finite density of state near the Fermi level, which is localized and
does not contribute to conduction. In this scenario, electronic
specific heat remains finite despite the system showing non-metallic
behavior. This is precisely in agreement with our observation in
Region II. Therefore, the non-metallic behavior below
$T_{\mathrm{MIT}}$ should primarily be ascribed to Anderson
localization.

As the magnetic state switches from the AFM state to the nearly FM
state near $x$ = 0.4, the electronic state changes to a weakly
localized state. Our analyses show that it results from magnetic scattering. As shown
in Fig. \ref{fig:RTB}a, the small upturn of resistivity at low
temperature for the $x = 0.3$ sample is gradually suppressed by magnetic
field and eventually tuned to a quasi quadratic temperature
dependence when the applied field is above 4 T. Similar suppression
of the more significant upturn of resistivity in the $x = 0.38$
sample is also observed (see the inset to Fig. 3a). These
results strongly support the assertion that the weak localization
behavior observed for Region III is due to magnetic scattering. In
addition, we note that the minimum resistivity occurs close to the
CSG phase freezing temperature $T_{\mathrm{f}}$. Similar behavior
has been observed in many other $d$- and $f$- electron spin glasses,
such as NiMn \cite{NiMn}, NiMnPt \cite{NiMnPt}, FeAl$_{2}$
\cite{FeAl2} and U$_{2}$PdSi$_{3}$ \cite{U2PdSi3}, and it is
considered to be associated with the formation of remanent FM
domains. For our case, FM correlations are progressively enhanced
with lowering temperature, eventually freezing into a CSG phase
\cite{SrCa327HMNF,SrCa327JPSJ}. In the CSG phase the FM clusters are
randomly frozen, resulting in strong magnetic scattering among FM
clusters. Under the application of magnetic field such magnetic
scattering is suppressed, thus suppressing the small upturn of
resistivity at low temperatures.

With decreasing Ca content, $T_{\mathrm{f}}$ decreases and FM
magnetic correlations become weaker \cite{SrCa327HMNF,SrCa327JPSJ},
thus reducing magnetic scattering. Consequently, the weakly
localized behavior gradually disappears, eventually yielding to a
fully metallic state in Region IV. The temperature dependence of
resistivity in this metallic state can be fit to
$\rho=\rho_{0}+AT^{\alpha}$ at low temperatures with $\alpha> 2$.
$\alpha$ is $~ 2.7$ for the $x = 0.2$ sample; it decreases with a
further decrease of $x$. When $x = 0.1$, $\alpha$ is decreased to
2.2, suggesting that the magnetic scattering is still involved in
the transport process in this sample. Thus, the electronic state in
Region IV can be viewed as a magnetic metallic state. When $x$ is
decreased below 0.08, the system enters a metamagnetic state and its
electronic state behaves as a Fermi liquid ground state. As seen in
Fig. \ref{fig:FL}, starting from $x = 0$, the Fermi liquid
temperature $T_{\mathrm{FL}}$ gradually decreases with increasing
\textit{x}, from 10 K for $x = 0$ down to zero for $x \approx 0.08$.
We observed a non-Fermi liquid behavior in the temperature
dependence of the electronic specific heat near the critical
composition ($x \approx 0.08$) in our early work \cite{SrCa327HMNF}.
Such a non-Fermi liquid behavior could be attributed to the system
approaching a magnetic instability near $x \sim 0.08$.

\begin{figure}
\includegraphics[angle=-90]{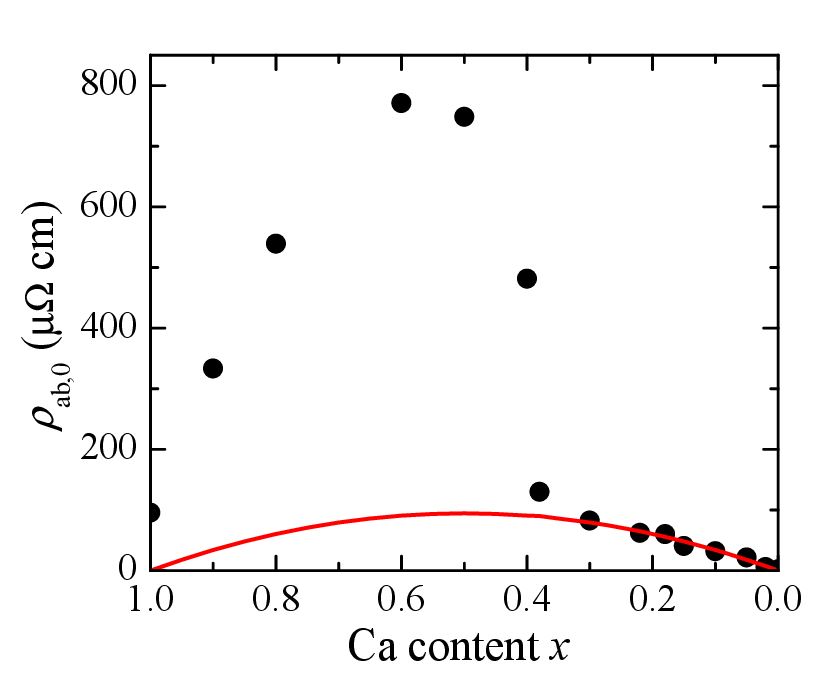}
\caption{(color online) Residual in-plane resistivity as a function
of the Ca content $x$. For those samples which show localized
behaviors at low temperatures, $\rho_{ab}$ at 0.3 K is taken as
the residual in-plane resistivity. The solid curve represents a fit
to the Nordheim law $Ax(1-x)$.}\label{fig:resR}
\end{figure}

Although magnetic scattering plays a critical role in transport
properties in region III and IV , it is still much weaker in
comparison with the scattering from the disorder introduced by Ca
substation for Sr. This can be clearly seen by fitting residual
resistivity to the Nordheim law, which is solely based on
disorder-induced scattering. As shown in Fig. \ref{fig:resR}, the
residual in-plane resistivity for $x < 0.4$ can approximately be
fitted to the Nordheim formula $Ax(1-x)$ with $A = 378$
$\mu\Omega\cdot \textrm{cm}$, implying that the randomness intrinsic
to the Sr/Ca substitution alone can account for the variation of
residual resistivity with the Ca content. Only when the electronic
state transitions to the Anderson localized state for $x \geq 0.4$,
show the residual resistivity (taken as the resistivity at 0.3 K),
significant deviation from this formula. Another reason for this
deviation is that the spin-valve effect sets in for $x\geq 0.4$,
which could dramatically affect the residual resistivity
\cite{Ca327elastic}.

Comparing the double layered ruthenates
(Sr$_{1-x}$Ca$_{x}$)$_{3}$Ru$_{2}$O$_{7}$ to their single layered
analogue Ca$_{2-x}$Sr$_{x}$RuO$_{4}$
\cite{SrCa214NakatsujiPRL1,SrCa214NakatsujiPRL2} we note that while
they undergo a similar nearly FM-to-AFM transition, distinct
differences exist between their electronic states. First, we have
shown that the electronic ground state near the Ca side in the
double layered ruthenates is an Anderson localized state, whereas
the insulating state near the Ca side in single layered ruthenates
is a Mott insulating state \cite{Ca214Mott,SrCa214NakatsujiPRL1}.
Second, while both solid solutions show CSG phases
\cite{SrCa327HMNF,SrCa214NakatsujiPRL2,SrCa327JPSJ}, the electronic
ground state differs significantly between them: a Fermi liquid
ground state is observed in the CSG phase of
Ca$_{2-x}$Sr$_{x}$RuO$_{4}$ \cite{SrCa214NakatsujiPRL2}, but in the
CSG phase of (Sr$_{1-x}$Ca$_{x}$)$_{3}$Ru$_{2}$O$_{7}$ the magnetic
scattering plays a critical role, resulting in a weakly localized
state.

Variations of electronic and magnetic properties in
(Sr$_{1-x}$Ca$_{x}$)$_{3}$Ru$_{2}$O$_{7}$ can all be attributed to
structure changes caused by Ca substitution for Sr. Since Ca$^{2+}$
is smaller than Sr$^{2+}$ in ionic radius, Ca$^{2+}$ substitution
for Sr$^{2+}$ should increase structural distortion as in
Ca$_{2-x}$Sr$_x$RuO$_4$ \cite{StructureCaSr214}. Consistent with
this expectation, Iwata \textit{et al.} \cite{SrCa327JPSJ} observed
discontinuous changes in lattice parameters near the boundary
between the AFM and the nearly FM phases, and bifurcation of the
$a$-axis lattice parameter for 0.6 $< x <$ 1. From our X-ray
diffraction measurements, we observed similar results. Furthermore,
we recently performed structure refinement of X-ray diffraction
spectra for most of the samples presented in the phase diagram in
Fig. \ref{fig:ps}. We found that the structure distortion caused by
the substitution occurs via rotation and tilting of RuO$_6$
octahedra \cite{SrCa327StrM}. Rotation angle gradually increases
with Ca substitution for Sr and approaches saturation for $x > 0.6$,
whereas the tilting does not occur until the Ca content $x$ is increased
above 0.4, and it enhances significantly for $x > 0.6$. These
results explain the magnetic phase transitions observed in this
system \cite{SrCa327StrM}, the enhanced in-plane magnetic anisotropy
for $x
> 0.6$ \cite{SrCa327StrM}, as well as the lattice parameter changes
observed by Iwata \textit{et al.} \cite{SrCa327JPSJ}. This, together
with our observation of complex electronic states described above,
suggests strong interplay between charge, spin, and lattice degrees
of freedom in double-layered ruthenates.

\section{Conclusion}

In summary, we have established an electronic phase diagram for
double layered ruthenates (Sr$_{1-x}$Ca$_{x}$)$_{3}$Ru$_{2}$O$_{7}$.
Our results show that the electronic states of this system are
complex and strongly coupled with the magnetic states. Disorder has
a remarkable effect on electronic transport properties. The
electronic ground state in Ca$_{3}$Ru$_{2}$O$_{7}$ is quasi-2D
metal, but transforms into an Anderson localized state in the AFM
region with 0.4$\leq x <$ 1.0 due to the presence of disorders
introduced by Sr substitution for Ca. When the magnetic state
switches from the AFM to the nearly FM state near $x$ = 0.4, the
electronic state changes to a weakly localized state  induced by
magnetic scattering for 0.25 $\leq x < $0.4, and then evolves into a
magnetic metallic state for 0.08 $< x <$ 0.25. When $x$ is decreased
below 0.08, the system enters a metamagnetic Fermi liquid ground
state. The Fermi liquid temperature is suppressed to zero near the
critical composition with $x \sim$ 0.08 and non-Fermi liquid
behavior occurs accordingly. Such complex electronic states coupled
with magnetic states testify the strong interplay of charges and
spin degrees of freedom in double-layered ruthenates.

\begin{acknowledgments}
We would like to acknowledge valuable discussions with Dr. I.
Vekhter and Dr. V. Dobrosavljevic. Work at Tulane is supported by
NSF under grant DMR-0645305 for materials and equipments, DOE under
DE-FG02-07ER46358 for a postdoc, the DOD ARO under Grant No.
W911NF-08-C0131 for students, and the Research Corporation, work at
UNO by DARPA under grant HR0011-07-1-0031.
\end{acknowledgments}

\end{document}